\documentclass[11pt]{article}
\usepackage[margin=1in]{geometry}
\usepackage[T1]{fontenc}
\usepackage[utf8]{inputenc}
\usepackage{lmodern}
\usepackage{microtype}
\usepackage{amsmath,amssymb}
\usepackage{booktabs}
\usepackage{graphicx}
\usepackage{caption}
\usepackage[round]{natbib}
\usepackage{placeins}
\usepackage[hidelinks]{hyperref}
\newcommand{\GateKoRet}{0.90}
\newcommand{\GateTier}{0.90}
\newcommand{\GateDacc}{-0.03}
\newcommand{\GateFmtPP}{2}
\newcommand{\GateMem}{5.98}
\newcommand{\GateTpot}{1.1}
\newcommand{\ZA}{2.2414}
\newcommand{\ZB}{2.5758}
\newcommand{\KA}{2}
\newcommand{\KB}{5}
\newcommand{\NFin}{5,168}
\newcommand{\NFinKo}{1,565}
\newcommand{\NFinEn}{1,043}
\newcommand{\NFinNum}{1,320}
\newcommand{\NFinRea}{870}

\newcommand{\NSel}{5,052}

\newcommand{\NDiag}{1,998}
\newcommand{\NFinGate}{4,798}
\newcommand{\NFinGen}{2,190}

\newcommand{\TMem}{15.27}

\newcommand{\TTtft}{15.3}
\newcommand{\TTok}{255.4}
\newcommand{\FNvMem}{5.917}

\newcommand{\FNvTier}{0.931}
\newcommand{\FNvKoRet}{0.904}
\newcommand{\FNvDacc}{$-$0.31}
\newcommand{\FNvTpot}{0.939}
\newcommand{\FNvTtft}{15.4}
\newcommand{\FNvTok}{272}

\newcommand{\FNfTier}{0.887}
\newcommand{\FNfKoRet}{0.821}

\newcommand{\FNfTtft}{42.3}
\newcommand{\FNfTok}{387}

\newcommand{\FNvMemBytes}{6,353,372,672}
\newcommand{\FNvKoMargin}{0.0036}
\newcommand{\FNvMemMargin}{0.063}
\newcommand{\FNvFmt}{+0.14}

\newcommand{\FNvNumDacc}{$-$0.91}
\newcommand{\FNvReaDacc}{$-$0.11}
\newcommand{\FNfReaDacc}{$-$3.91}
\newcommand{\FNvSup}{+1.4}
\newcommand{\FNvSupCI}{[+0.2, +2.6]}
\newcommand{\FNvSupLo}{+0.2}
\newcommand{\FNvSupKo}{+1.1}
\newcommand{\FNvSupKoCI}{[$-$1.1, +3.3]}
\newcommand{\FNvSupEn}{+1.2}
\newcommand{\FNvSupEnCI}{[$-$0.3, +2.7]}
\newcommand{\FNvSupNum}{+0.4}
\newcommand{\FNvSupNumCI}{[$-$1.9, +2.6]}
\newcommand{\FNvSupRea}{+3.8}
\newcommand{\FNvSupReaCI}{[+0.3, +7.2]}
\newcommand{\FBjSup}{+4.1}
\newcommand{\FBjSupCI}{[+2.3, +5.9]}
\newcommand{\FinVerdict}{D81\_DEPLOY\_CANDIDATE}
\newcommand{\SNvTier}{0.935}
\newcommand{\SNvKoRet}{0.924}
\newcommand{\SNvMem}{5.917}
\newcommand{\SNvSup}{+3.4}
\newcommand{\SNvSupCI}{[+2.3, +4.6]}

\newcommand{\SWwKoRet}{0.869}

\newcommand{\SWwSup}{$-$0.5}
\newcommand{\SWwSupCI}{[$-$1.7, +0.6]}
\newcommand{\SBjTier}{0.808}
\newcommand{\SBjKoRet}{0.726}

\newcommand{\SBjSup}{+5.1}
\newcommand{\SBjSupCI}{[+3.3, +7.0]}

\newcommand{\SNvTpot}{0.925}

\newcommand{\CalKo}{+2.2}
\newcommand{\CalKoCI}{[$-$0.9, +5.4]}

\newcommand{\CalRea}{$-$12.3}
\newcommand{\CalReaCI}{[$-$16.8, $-$7.9]}

\newcommand{\AlNvUnits}{46}
\newcommand{\AlNvUsed}{0.499}
\newcommand{\AlNvBudget}{0.50}
\newcommand{\AlWwUnits}{53}

\newcommand{\AlBjUnits}{44}
\newcommand{\AlBjUsed}{0.251}
\newcommand{\AlBjBudget}{0.80}
\newcommand{\AlBjBudgetA}{0.30}
\newcommand{\AlBjBudgetB}{0.55}
\newcommand{\AlNvQkv}{23}
\newcommand{\AlNvO}{18}
\newcommand{\AlNvGu}{1}
\newcommand{\AlNvDown}{4}

\newcommand{\AlNvPred}{5.947}
\newcommand{\AlNvLimit}{5.949}
\newcommand{\AlNvHead}{2.01}
\newcommand{\MemNvaSixteen}{6.009}
\newcommand{\MemNvaLmh}{5.44}
\newcommand{\MemLmhSave}{0.57}

\newcommand{\DvHmTier}{0.926}
\newcommand{\DvHmKoRet}{0.908}

\newcommand{\FeHmTier}{0.891}
\newcommand{\FeHmKoRet}{0.877}

\newcommand{\DvBmMem}{5.191}
\newcommand{\FeBmTier}{0.804}
\newcommand{\FeBmKoRet}{0.730}

\newcommand{\DvNvaVsNf}{+2.8}
\newcommand{\DvNvaVsNfCI}{[+1.6, +4.0]}
\newcommand{\DvNvhVsNva}{$-$0.3}
\newcommand{\DvNvhVsNvaCI}{[$-$1.3, +0.8]}
\newcommand{\DvHmVsWh}{+1.6}
\newcommand{\DvHmVsWhCI}{[+0.4, +2.8]}
\newcommand{\DvWhVsWf}{$-$0.6}
\newcommand{\DvWhVsWfCI}{[$-$1.7, +0.6]}

\newcommand{\FeBmVsWt}{+10.3}
\newcommand{\FeBmVsWtCI}{[+8.6, +12.0]}

\newcommand{\FeHmVsWf}{$-$2.4}
\newcommand{\FeHmVsWfCI}{[$-$3.6, $-$1.2]}
\newcommand{\FeHmRea}{$-$9.0}
\newcommand{\DvHmRea}{$-$5.1}
\newcommand{\DsWtTier}{0.748}
\newcommand{\DsWtMem}{4.86}

\newcommand{\DsWtwoMem}{4.07}

\newcommand{\DsWtVsRt}{+29.9}
\newcommand{\DsWtVsRtCI}{[+27.9, +32.0]}

\newcommand{\DsTwoRed}{73.3}
\newcommand{\DsThreeRed}{68.2}
\newcommand{\WtwoBest}{0.243}
\newcommand{\WtwoGpuH}{4.49}
\newcommand{\DgCalKo}{+12.2}
\newcommand{\DgCalKoCI}{[+2.3, +22.1]}
\newcommand{\DgCalRea}{+7.4}
\newcommand{\GpuFin}{1.35}
\newcommand{\GpuSel}{2.39}
\newcommand{\GpuDiag}{4.98}
\newcommand{\FinSha}{7e6fcc531386}
\newcommand{\SelSha}{9c4bce0c5e91}
\newcommand{\NEightyFin}{4,061}
\newcommand{\NEightyDev}{3,003}
\newcommand{\NEightyDiag}{1,662}
\newcommand{\FeBmVsWtKo}{+5.8}

\newcommand{\FeBmVsWtNum}{+19.4}
\newcommand{\FeBmVsWtRea}{+12.4}
\newcommand{\FNvKernNv}{98}
\newcommand{\FNvKernMarlin}{46}
\newcommand{\FBjKernHum}{100}
\newcommand{\FBjKernMarlin}{44}

\newcommand{\FmBfTrunc}{4.4}
\newcommand{\FmBfNone}{4.5}
\newcommand{\FmBfLoop}{2.3}

\newcommand{\FmRtTrunc}{54.4}

\newcommand{\FmRtLoop}{56.0}

\newcommand{\FmWtwoTrunc}{99.9}
\newcommand{\FmWtwoNone}{100.0}
\newcommand{\FmWtwoLoop}{100.0}

\newcommand{\FmGenN}{1,302}
\newcommand{\AlHmUnits}{24}

\newcommand{\AlBmUnits}{41}

\newcommand{\AlBmPred}{5.159}

\newcommand{\NvCkptBytes}{6,320,384,652}
\newcommand{\FinSetSha}{26a1fadc91b6}

\date{2026-11-16}   

\title{Beyond Bitwidth: Deployment-Constrained Mixed-Precision Quantization\\
for Multilingual LLM Inference on NVIDIA Blackwell}
\author{Hyojung Han\\[4pt]
        \small ThakiCloud\\[2pt]
        \small \texttt{hyojung.han@thakicloud.com}}

\begin{document}
\maketitle

\begin{abstract}
Low-bit LLM quantization is useful only if a model preserves service-relevant accuracy, executes through
supported GPU kernels, and meets measured memory and latency limits. We study Qwen3-8B on NVIDIA B200 with
vLLM, including INT2/INT3, GPTQ, NVFP4, and mixed-precision configurations. We verify native packed
execution and precision on all 144 fused linear modules and evaluate Korean, English, numerical, and
reasoning quality under preregistered gates. Packed INT2 and INT3 execute, but all seven tested INT2 grids
degenerate on generation. Isolated module-restoration effects are strongly non-additive; Korean and
reasoning sensitivity rankings disagree, and a promising INT4/INT8 development candidate fails independent
final evaluation. A configuration with 16-bit activations, 98 NVFP4 weight-only modules, 46 INT8 modules,
and an INT8 output head passes all gates on the untouched FINAL-81 set (\NFin{} items): \FNvMem{}~GiB of
weight-loading memory (limit \GateMem{}), Korean teacher-correct retention \FNvKoRet{} (gate
\GateKoRet{}), and TPOT ratio \FNvTpot{} relative to BF16. It improves paired accuracy over NVFP4 W4A4 by
\FNvSup{} percentage points (Bonferroni-adjusted interval \FNvSupCI{}), compared with \SNvSup{} points on
the selection set. This comparison changes both activation precision and INT8 protection; exploratory
analyses suggest that the activation change contributes substantially, but the isolated allocation effect
is not established. Final checkpoints were not retained, and bitwise rebuild reproducibility remains
unverified.
\end{abstract}

\section{Introduction}

Low-bit quantization is the standard way to cut the memory footprint and serving cost of large language
models. GPTQ \citep{frantar2023gptq} and AWQ \citep{lin2024awq} showed that weight precision can be reduced
substantially while preserving most of a model's capability, and mixed-precision methods
\citep{zheng2026mixllm,zhao2025impq} assign different precisions to different components to improve the
accuracy--compression trade-off. GPU hardware has meanwhile added native support for microscaling formats
such as NVFP4, although accuracy and speed depend strongly on the format, its numerical details and the
runtime implementation \citep{egiazarian2026fp4}.

A gap therefore opens between quantization feasibility and deployment feasibility. A compact checkpoint can
still need more GPU memory than expected once loaded. A configuration can run through an optimized packed
kernel and still produce unusable text. A format can deliver high output-token throughput while the first
token arrives late, or while one language degrades far more than another. Language-aware calibration has
been studied \citep{chimoto2026calibrating,waskito2026multilinguality}; here we evaluate its task-specific
effects alongside measured runtime memory and kernel compatibility constraints.

We study these interactions on Qwen3-8B \citep{qwen3} served by vLLM \citep{vllm029} on an NVIDIA B200
GPU, through three preregistered experiment stages (D79, D80 and D81). Instead of optimizing a single
compression ratio or benchmark score, each candidate must jointly satisfy four conditions: it runs through
the intended packed kernel for every fused module; its measured weight-loading memory stays below a fixed
limit; its Korean, English, numerical and reasoning quality stays within preregistered tolerances of the
BF16 teacher; and its time per output token (TPOT) does not exceed that of BF16 by more than 10\%.

One instructive result is a failure. In D80, a mixed INT4/INT8 configuration ranked first on
the development set and then failed the quality gates on the independent final set. D81 was designed
around that lesson, with separate diagnostic, selection and final sets and a rule that no research-only
candidate may be promoted to a deployment candidate on the strength of a final score. Under that protocol
one configuration, which combines NVFP4 weight-only quantization with INT8 protection, passed every gate on
both the selection set and the untouched final set.

\paragraph{Contributions.}
\begin{itemize}
  \item \textbf{An end-to-end validation workflow} that checks packed execution and weight type per fused
        module, measured weight-loading memory, per-axis quality against a teacher, and serving latency,
        instead of relying on nominal bitwidth (Section~\ref{sec:method}).
  \item \textbf{A characterization of low-bit failure modes} on B200: INT2 and INT3 execute natively, yet
        INT2 degenerates on generation under every grid we tried and INT3 leaves large Korean and
        reasoning losses (Sections~\ref{sec:lowbit}--\ref{sec:mix34}).
  \item \textbf{Evidence against isolated sensitivity heuristics}: restoration effects are strongly
        non-additive, Korean and reasoning rankings largely disagree, and a development-set winner did not
        survive held-out evaluation (Sections~\ref{sec:sens}--\ref{sec:devfinal}).
  \item \textbf{A deployment-constrained configuration confirmed on a held-out set}, together with
        exploratory evidence that activation precision is a major contributor to its advantage; the
        independent effect of the INT8 allocation remains unidentified (Sections~\ref{sec:d81}--\ref{sec:attr}).
\end{itemize}
We did not invent GPTQ, NVFP4, mixed-precision quantization or any kernel. The contribution is an
empirical study under deployment constraints and a configuration-selection protocol whose failures we
report alongside its successes.

\section{Background and related work}

\paragraph{Weight-only quantization and rounding.}
Early deployment-oriented quantization work established the importance of exceptional activations,
per-channel scaling, and hardware-aware compression. \citet{dettmers2022llmint8} isolate outliers in 8-bit
transformer inference, while \citet{yao2022zeroquant} co-design a quantization pipeline and its inference
backend. GPTQ \citep{frantar2023gptq} uses second-order error compensation for weight-only quantization, and
AWQ \citep{lin2024awq} exploits activation statistics to protect salient weights. Nontrivial rounding and
clipping choices are studied by AdaRound \citep{nagel2020adaround} and OmniQuant \citep{shao2024omniquant}.
SpQR \citep{dettmers2024spqr} and SqueezeLLM \citep{kim2024squeezellm} protect difficult weights using sparse
or mixed representations. We use established GPTQ and RTN recipes; we neither modify their optimizers nor
compare them directly with the full range of these alternatives. Our emphasis is on whether the resulting
configuration is executable, satisfies a measured memory ceiling, and preserves preregistered task
capabilities.

\paragraph{Extreme low-bit representations.}
QuIP \citep{chee2023quip} and QuIP\# \citep{tseng2024quipsharp} combine incoherence processing with adaptive
rounding and, in the latter, lattice codebooks. AQLM \citep{egiazarian2024aqlm} uses additive codebook
quantization. ParetoQ \citep{liu2025paretoq} studies training and scaling across extremely low bitwidths.
Quantization-aware training also offers a different recovery path: LLM-QAT \citep{liu2024llmqat} and
EfficientQAT \citep{chen2025efficientqat} use training or layer-wise optimization to reduce quantization
damage. QLoRA \citep{dettmers2023qlora} addresses efficient adaptation of models with quantized frozen
weights, a distinct setting from our post-training inference experiments. These methods are relevant
alternatives to our native INT2/INT3 diagnostics; their reported success prevents generalizing our INT2
failures into a claim that two-bit language models are inherently unusable. We did not implement or
benchmark these alternatives on our fixed B200 runtime.

\paragraph{Activation precision and FP4.}
SmoothQuant \citep{xiao2023smoothquant} transfers quantization difficulty from activations to weights.
QuaRot \citep{ashkboos2024quarot}, SpinQuant \citep{liu2025spinquant}, DuQuant \citep{lin2024duquant}, and
FlatQuant \citep{sun2025flatquant} employ different transformations or rotation strategies to control
activation outliers. Recent microscaling FP4 work \citep{egiazarian2026fp4} directly examines the gap
between the accuracy and hardware promise of FP4 formats. These methods make an important distinction for
our study: NVFP4 W4A4 and NVFP4 W4A16 are not equivalent comparisons, because they use different activation
precision. Our FINAL-81 \texttt{A81} versus W4A4 result changes both activation precision and weight
protection; it is therefore a configuration-level comparison, not an identified allocation effect.

\paragraph{Mixed-precision allocation and interaction effects.}
Classical mixed-precision methods such as HAWQ \citep{dong2019hawq} and HAWQ-V2 \citep{dong2020hawqv2} use
sensitivity information to help allocate precision, including second-order and Pareto-based approaches.
Atom \citep{zhao2024atom} co-designs low-bit quantization and runtime throughput. More recent LLM
approaches include MixLLM \citep{zheng2026mixllm}, which mixes precision across output features, and IMPQ
\citep{zhao2025impq}, which considers interactions between layers. Our non-additive restoration results
(Section~\ref{sec:sens}) motivate such interaction-aware methods, but our frozen allocation procedure is a
greedy heuristic, not an interaction-aware optimizer. Without matched-memory comparisons to these methods
we make no methodological superiority claim.

\paragraph{GPU serving and memory accounting.}
The PagedAttention/vLLM system \citep{kwon2023pagedattention} addresses dynamic KV-cache memory; Marlin
\citep{frantar2024marlin} optimizes weight-only quantized linear inference, and FlashInfer
\citep{ye2025flashinfer} studies customizable GPU serving kernels. QServe \citep{lin2025qserve} explicitly
co-designs low-bit quantization and serving, whereas KIVI \citep{liu2024kivi} targets KV-cache precision
rather than model-weight precision. We separately measure the vLLM weight-loading allocation increment and
token latency. We do not treat the former as total peak serving memory, nor report KV-cache or concurrency
improvements that were not measured. The Humming kernels \citep{humming2026} provide the native INT2/INT3
execution path in our pinned vLLM environment \citep{vllm029}; we implement no matrix-multiplication
kernel of our own.

\paragraph{Multilingual calibration and evaluation.}
\citet{chimoto2026calibrating} investigate multilingual calibration for GPTQ and AWQ, while
\citet{waskito2026multilinguality} examine calibration language, bitwidth and cross-lingual effects. These
studies motivate a per-language rather than solely aggregate quality analysis. Our Korean-enriched
calibration experiment (Section~\ref{sec:cal}) is a configuration-specific counterexample to assuming that
adding more target-language calibration text necessarily improves target-language performance without
harming reasoning. The observation does not contradict benefits reported in other models or calibration
settings.

\section{Deployment-constrained quantization}
\label{sec:method}

\subsection{Model and quantization units}
Qwen3-8B \citep{qwen3} has 36 transformer layers. vLLM executes each layer through four fused linear
modules: the fused query--key--value projection (\texttt{qkv}), the output projection (\texttt{o}), the
fused gate and up projection (\texttt{gate\_up}) and the down projection (\texttt{down}), giving 144 fused
modules in total. We assign precision at these boundaries because the runtime requires one scheme per fused
module. In a negative control, assigning W4 to the query projection and W8 to the key and value projections
inside one fused \texttt{qkv} module made the vLLM weight loader reject the checkpoint with a shape
assertion. All configurations below therefore use one precision per fused module. The output head
(\texttt{lm\_head}) is treated separately.

\subsection{Formats and runtime paths}
Table~\ref{tab:formats} lists the families we evaluate and the kernels vLLM 0.29.0 selected for them on
B200. Its WNA16 path accepts symmetric integer weights from 2 to 8 bits and asymmetric weights only at 4
and 8 bits. Kernel names describe the measured configuration, not a claim that a kernel family is
generally optimal.

\begin{table}[t]
\centering
\caption{Quantization families and the kernels vLLM 0.29.0 selected on B200.}
\label{tab:formats}
\begin{tabular}{llll}
\toprule
Family & Weights & Activations & Kernel \\
\midrule
BF16 & BF16 & BF16 & reference \\
NVFP4 W4A4 (n4) & NVFP4 & FP4 & FlashInferCuteDslNvFp4 \\
NVFP4 W4A16 & NVFP4 & 16-bit & MarlinNvFp4 \\
GPTQ W8A16 / W4A16 & INT8 / INT4 & 16-bit & Marlin (uint8b128 / uint4b8) \\
GPTQ W3A16 & INT3 & 16-bit & Humming (uint3b4) \\
GPTQ W2A16 & INT2 & 16-bit & Humming (uint2b2) \\
\bottomrule
\end{tabular}
\end{table}

\subsection{Memory constraint}
Checkpoint size, the sum of registered parameter bytes and the GPU memory allocated while loading can all
differ. Our memory gate uses vLLM's raw weight-loading counter, the change in allocated GPU memory while
the model loads:
\begin{equation}
  M(c) = \texttt{weight\_load\_bytes}(c) / 2^{30} \;<\; \GateMem~\text{GiB},
\end{equation}
with strict inequality. The limit is a fixed product requirement of this study. During allocation we
predict $M(c)$ from earlier measurements with a 32~MiB safety margin, but a prediction never counts as
passing: each candidate is first loaded in a smoke job that does not read the selection set, and only a
measured counter below the limit admits it to selection. The counter excludes KV cache and runtime
activations.

\subsection{Quality and latency constraints}
Let $T$ be the BF16 teacher and $Q$ a quantized configuration. For an evaluation axis $a$ we measure
teacher-correct retention and the change in accuracy,
\begin{equation}
  r_a(Q) = P\!\left(Q~\text{correct} \mid T~\text{correct}\right), \qquad
  \Delta_a(Q) = \mathrm{acc}_a(Q) - \mathrm{acc}_a(T).
\end{equation}
Retention measures how many of the teacher's correct decisions survive; the accuracy change also credits
items the teacher gets wrong and the quantized model gets right. The tier is retention pooled over all items
of the four gate axes (Korean, English, numerical, reasoning). A configuration passes the quality gates if
tier $\ge \GateTier$, $r_{\mathrm{ko}} \ge \GateKoRet$, overall, numerical and reasoning accuracy changes
are each at least $\GateDacc$ (three points), and the rate of malformed outputs on generation items rises
by at most \GateFmtPP{} points. It passes serving if $M(c) < \GateMem$~GiB and its median TPOT is at most
\GateTpot{} times that of BF16. Execution must match the expected packed kernel for every fused module,
with no layer silently expanded to floating point, and the software environment must match the frozen
manifest. Time to first token (TTFT) is reported but not gated. These thresholds were fixed before D78 and
not changed in any later stage.

\subsection{Execution verification}
Kernel names alone are not enough: INT4 and INT8 both run through Marlin, so the stored weight type is
checked separately, and an NVFP4 scheme name does not reveal whether activations are quantized. For every
candidate the checker records the kernel and weight type of each fused module, the activation policy, the
output head, the absence of floating-point fallbacks, the model revision and the raw memory counter.

\subsection{Sensitivity diagnostics and allocation}
Diagnostic runs, on a set reserved for them, measure how much each fused module contributes: starting from
a fully quantized base, one module is restored to BF16 at a time and the change in teacher-correct
retention is recorded per axis. We use these measurements to rank modules, not to predict the quality of a
combination. A frozen script orders modules by their joint restoration effect (or a Korean-weighted variant)
and a second frozen script fills the memory budget from the top of the list, stopping at the first module
that does not fit. Both scripts and their output hashes were fixed before the selection set was read.
Restoration is a proxy: the deployed protection format is INT8, not BF16. We do not claim that this greedy
procedure is optimal.

\section{Experimental protocol}

\subsection{Hardware and software}
All runs use one NVIDIA B200 with vLLM 0.29.0, PyTorch 2.11.0 and transformers 5.16.1. Checkpoints are
produced with llm-compressor 0.15.0 and compressed-tensors 0.20.1 from a pinned Qwen3-8B revision. Each
job verifies its environment against the stage's frozen manifest before evaluating. Serving latency is
measured at batch size 1 over 32 prompts with 256 forced output tokens. Measured GPU time for the D81 stages
was \GpuDiag{} (diagnostic), \GpuSel{} (selection) and \GpuFin{} (final) B200 GPU-hours, excluding queueing.

\subsection{Evaluation sets}
Items come from KMMLU \citep{son2025kmmlu}, KoBEST \citep{jang2022kobest}, MMLU \citep{hendrycks2021mmlu},
ARC \citep{clark2018arc}, HellaSwag \citep{zellers2019hellaswag}, BIG-Bench Hard \citep{suzgun2023bbh},
BIG-bench \citep{srivastava2023bigbench}, MuSR \citep{sprague2024musr}, HRM8K \citep{ko2025hrm8k} and a
seeded generator of numerical problems. Multiple-choice items are scored by log-likelihood; generation items
use greedy decoding with output limits of 1,536 tokens (numerical), 4,096 (reasoning) and 768 (Korean math).
Each stage builds its sets with a script that removes every item used in earlier stages, so no item is
reused across D78--D81.

D80 used diagnostic, development and final sets of \NEightyDiag{}, \NEightyDev{} and \NEightyFin{} items.
D81 used a diagnostic set of \NDiag{} items, a selection set (SEL-81) of \NSel{} items and a final set
(FINAL-81) of \NFin{} items. FINAL-81 contains \NFinGate{} gate-axis items (Korean \NFinKo{}, English
\NFinEn{}, numerical \NFinNum{}, reasoning \NFinRea{}), of which \NFinGen{} are generation items, plus 370
report-only items, 300 of them HRM8K Korean math. No evaluation item is reproduced in this paper. Dataset
access and redistribution rights are handled separately from the manuscript and must be checked against each
source's terms before any evaluation-set release.

\subsection{Preregistration and statistics}
Candidate rules, gates, comparison families and the selection rule were frozen and committed before each
set was read. Paired differences use identical items. With $C_Q(i)\in\{0,1\}$ the correctness of $Q$ on
item $i$, the primary metric is
\begin{equation}
  \Delta(Q, R) = \frac{1}{n}\sum_{i=1}^{n} \left[C_Q(i) - C_R(i)\right]
\end{equation}
over all gate-axis items, with a normal-approximation interval. D79 and D80 report 95\% intervals. In D81
each track is a Bonferroni family: track A (four-bit-class candidates) has $k=\KA$ and $z=\ZA$, and
track B (three-bit-class candidates) has $k=\KB$ and $z=\ZB$. Superiority requires the adjusted lower bound
to exceed zero. Two track-B budget variants produced exactly the same allocation as a third
(Section~\ref{sec:d81}); they were not run and are not counted as observations, and $k=\KB$ was kept.
Deployment eligibility and superiority are assessed separately: a candidate can pass every absolute gate
without beating its reference, or beat its reference while failing the gates.

\subsection{Selection and final evaluation}
The diagnostic set builds candidates, the selection set chooses them under the frozen rule, and the final
set is read once for the chosen candidates and their references. A track with no candidate that passes all
gates has no deployment candidate; a research-only candidate is never promoted because it scores well on
the final set.

\section{Results}

\subsection{Native low-bit execution does not guarantee quality}
\label{sec:lowbit}

D79 tested whether very-low-bit Qwen3-8B checkpoints run on B200. Packed INT3 and INT2 checkpoints loaded
and executed through the Humming kernels for all 144 fused modules. Measured weight-loading memory fell to
\DsWtMem{}~GiB at W3 and \DsWtwoMem{}~GiB at W2, against \TMem{}~GiB for BF16, reductions of
\DsThreeRed\% and \DsTwoRed\% (Table~\ref{tab:d79}).

\begin{table}[t]
\centering
\caption{D79 development set: memory, effective bits per weight (BPW, including the output head and
embeddings), tier, Korean teacher-correct retention and TPOT ratio to BF16. Memory is the measured
weight-loading counter in GiB.}
\label{tab:d79}
\begin{tabular}{lrrrrr}
\toprule
Arm & Mem & BPW & Tier & $r_{\mathrm{ko}}$ & TPOT \\
\midrule
GPTQ W4 & 5.68 & 5.93 & 0.898 & 0.832 & 0.846 \\
GPTQ W3 & 4.86 & 5.08 & 0.748 & 0.575 & 0.857 \\
RTN W3 & 4.86 & 5.08 & 0.361 & 0.467 & 0.857 \\
GPTQ Attn W2 + MLP W4 (h2) & 5.33 & 5.56 & 0.521 & 0.459 & 0.864 \\
GPTQ Attn W4 + MLP W2 (h1) & 4.42 & 4.60 & 0.120 & 0.138 & 0.829 \\
GPTQ W2 & 4.07 & 4.23 & 0.115 & 0.158 & 0.837 \\
RTN W2 & 4.07 & 4.23 & 0.143 & 0.284 & 0.839 \\
\bottomrule
\end{tabular}

\end{table}

Quality is another matter. GPTQ improved on RTN at W3 by \DsWtVsRt{} points of paired accuracy
\DsWtVsRtCI{}, but GPTQ W3 remained far below the gates (tier \DsWtTier{}). The INT2 failure is qualitative.
On \FmGenN{} generation items, GPTQ W2 hit the output limit on \FmWtwoTrunc\% of items, produced no
extractable answer on \FmWtwoNone\% and repeated itself on \FmWtwoLoop\%, against \FmBfTrunc\%,
\FmBfNone\% and \FmBfLoop\% for BF16. RTN W3 already showed the same pattern in part (\FmRtTrunc\%
truncated, \FmRtLoop\% looping). Executing natively is necessary for deployment but tells us nothing about
whether the output is usable.

\subsection{Changing the INT2 grid does not stop degeneration}
\label{sec:int2}

We then asked whether the INT2 failure comes from the particular symmetric grid. Seven grids were
evaluated on the D80 diagnostic set, including symmetric, ternary, balanced four-level and asymmetric
variants (Table~\ref{tab:int2}). Asymmetric and balanced grids recovered some multiple-choice decisions,
the best reaching a retention of \WtwoBest, but every grid looped and truncated on all generation items
and retained none of the teacher's correct generations. This rules out the simplest explanation, that
changing the grid alone restores generation. It is not an impossibility result for INT2: we did not try
quantization-aware training or other models.

\begin{table}[t]
\centering
\caption{D80 INT2 grid diagnostics (diagnostic set, \WtwoGpuH{} GPU-hours). Retention is teacher-correct
retention over all gate items; MC and Gen.\ split it into multiple-choice and generation items. Loop and
Trunc.\ are the fractions of generation outputs that repeat or hit the output limit.}
\label{tab:int2}
\begin{tabular}{lrrrrr}
\toprule
Grid & Retention & MC & Gen. & Loop & Trunc. \\
\midrule
4-level balanced, MSE & 0.243 & 0.454 & 0.000 & 1.00 & 1.00 \\
GPTQ asym. INT2 & 0.209 & 0.392 & 0.000 & 1.00 & 1.00 \\
asym. INT2, MSE & 0.203 & 0.379 & 0.000 & 1.00 & 1.00 \\
GPTQ sym. INT2 & 0.163 & 0.304 & 0.000 & 1.00 & 1.00 \\
sym. INT2, min-max & 0.158 & 0.296 & 0.000 & 1.00 & 1.00 \\
sym. INT2, MSE & 0.151 & 0.283 & 0.000 & 1.00 & 1.00 \\
ternary, MSE & 0.120 & 0.225 & 0.000 & 1.00 & 1.00 \\
\bottomrule
\end{tabular}

\end{table}

\subsection{Mixed INT3/INT4 precision helps but stays below the gates}
\label{sec:mix34}

The D80 mixed configuration keeps \AlBmUnits{} fused modules in INT4 (Marlin) and the remaining 103 in
INT3 (Humming), at a measured \DvBmMem{}~GiB (the allocator had predicted \AlBmPred{}~GiB). On the D80 final
set it beat GPTQ W3 by \FeBmVsWt{} points \FeBmVsWtCI{}, with gains of \FeBmVsWtNum{} points on numerical
items, \FeBmVsWtRea{} on reasoning and \FeBmVsWtKo{} on Korean. Its Korean retention was still only
\FeBmKoRet{} and its tier \FeBmTier{} (Table~\ref{tab:d80}). A relative improvement over a weak baseline is
not deployment feasibility.

\begin{table}[t]
\centering
\caption{D80 arms on the development (dev) and final (fin) sets: memory (GiB), tier, Korean retention
and reasoning accuracy change relative to BF16 (points).}
\label{tab:d80}
\begin{tabular}{lrrrrrrr}
\toprule
Arm & Mem & Tier$_{\mathrm{dev}}$ & Tier$_{\mathrm{fin}}$ & $r_{\mathrm{ko,dev}}$ & $r_{\mathrm{ko,fin}}$ & $\Delta_{\mathrm{rea,dev}}$ & $\Delta_{\mathrm{rea,fin}}$ \\
\midrule
NVFP4 W4A4 (n4) & 5.977 & 0.894 & 0.879 & 0.838 & 0.805 & $-$3.9 & $-$3.9 \\
GPTQ W4 (w4) & 5.684 & 0.909 & 0.900 & 0.864 & 0.828 & $-$8.7 & +2.6 \\
W4 + W8 units + W8 head & 5.980 & 0.926 & 0.891 & 0.908 & 0.877 & $-$5.1 & $-$9.0 \\
GPTQ W3 (w3) & 4.864 & 0.753 & 0.667 & 0.634 & 0.578 & $-$23.8 & $-$17.7 \\
INT4/INT3 mix (D80) & 5.191 & 0.856 & 0.804 & 0.783 & 0.730 & $-$0.9 & $-$5.4 \\
\bottomrule
\end{tabular}

\end{table}

\subsection{Isolated sensitivity predicts combined restoration poorly}
\label{sec:sens}

The D81 diagnostics, run on five quantized bases, show two related effects (Table~\ref{tab:diag}). First,
Korean and reasoning sensitivities rank modules very differently: Spearman correlations between the two
rankings range from \(-0.046\) to 0.256. A single ranking cannot protect both capabilities. Second, effects
do not add up. On the two track-A bases, summing the Korean retention gains from restoring each module
alone gives 5.4 to 7.6 times the gain from restoring all of them together; on the INT3 bases the ratio is
negative. A sensitivity ranking can nominate candidates, but the quality of the resulting mixture has to be
measured directly.

\begin{table}[t]
\centering
\caption{D81 diagnostics. $\rho_{\mathrm{S}}$: Spearman correlation between the Korean and the reasoning
restoration rankings over the 144 fused modules. Last column: sum of single-module Korean retention gains
divided by the gain from restoring all modules jointly. Track-B bases differ only in calibration
(0\%, 25\% or 50\% Korean rows).}
\label{tab:diag}
\begin{tabular}{lrr}
\toprule
Base (track) & $\rho_{\mathrm{S}}(\mathrm{ko},\mathrm{rea})$ & $\sum$single / joint \\
\midrule
A: W4 body & 0.018 & 7.57 \\
A: NVFP4A16 body & 0.256 & 5.39 \\
B: W3, cal0 & $-$0.046 & $-$7.46 \\
B: W3, cal25 & 0.152 & $-$2.42 \\
B: W3, cal50 & $-$0.045 & $-$4.50 \\
\bottomrule
\end{tabular}

\end{table}

\subsection{A development-set gain disappeared on the final set}
\label{sec:devfinal}

The D80 INT4/INT8 configuration protects \AlHmUnits{} fused modules and the output head in INT8 and stores
the other 120 modules in INT4. On the development set it was the strongest four-bit-class arm, with tier
\DvHmTier{} and Korean retention \DvHmKoRet{}. On the final set its tier dropped to \FeHmTier{}, its Korean
retention to \FeHmKoRet{} and its reasoning change to \FeHmRea{} points (\DvHmRea{} on dev), so it failed
the gates (Table~\ref{tab:d80}). An exploratory comparison outside the preregistered family put it
\FeHmVsWf{} points \FeHmVsWfCI{} below plain GPTQ W4. Plausible causes include selection optimism, a shift
in the item mix and interactions among the protected modules; the data do not single one out. This result
is why D81 separates diagnostic, selection and final sets.

\subsection{D81: selection and final evaluation}
\label{sec:d81}

D81 built five candidates from the diagnostics under frozen rules. In track A, \texttt{A81} protects
\AlNvUnits{} fused modules (\AlNvQkv{} \texttt{qkv}, \AlNvO{} \texttt{o}, \AlNvGu{} \texttt{gate\_up},
\AlNvDown{} \texttt{down}) in INT8 on an NVFP4 W4A16 body with an INT8 output head, using \AlNvUsed{} of
its \AlNvBudget{}~GiB budget; its reference is NVFP4 W4A4 (n4). A parallel candidate on a GPTQ W4 body
(\AlWwUnits{} INT8 modules) has GPTQ W4 as its reference. In track B, \texttt{B81} upgrades \AlBjUnits{}
modules of a GPTQ W3 body to INT4; a second candidate uses a Korean-weighted ranking and a third repeats
\texttt{B81} with Korean-heavy calibration. The track-B allocator stopped at \AlBjUsed~GiB under every budget
it was given (\AlBjBudgetA, \AlBjBudgetB{} and \AlBjBudget~GiB), because the next modules in rank order were
excluded as harmful; the two smaller budgets therefore yielded the identical artifact and were treated as
aliases. Track B was preregistered as underpowered: the planned interval half-width needed about 6,100
items, more than SEL-81 provides, so non-significant track-B differences are not evidence of no effect.

On SEL-81, \texttt{A81} was the only candidate that passed every gate (tier \SNvTier{}, Korean retention
\SNvKoRet{}, \SNvMem{}~GiB, TPOT ratio \SNvTpot{}) and it beat n4 by \SNvSup{} points \SNvSupCI{}. The
GPTQ-W4-body candidate failed the Korean gate (\SWwKoRet) and did not beat GPTQ W4 (\SWwSup{} points
\SWwSupCI{}). \texttt{B81} beat GPTQ W3 by \SBjSup{} points \SBjSupCI{} but failed the quality gates
(tier \SBjTier{}, Korean retention \SBjKoRet{}), so track B had no deployment candidate and \texttt{B81}
went to the final set as a research comparison only.

\begin{table}[t]
\centering
\caption{FINAL-81 (\NFin{} items, read once). Mem: weight-loading memory (GiB); BPW: effective bits per
weight; Tier and $r_{\mathrm{ko}}$: teacher-correct retention; $\Delta$acc, $\Delta_{\mathrm{num}}$,
$\Delta_{\mathrm{rea}}$: accuracy change versus BF16 (points); $\Delta$fmt: change in the malformed-output
rate on generation items (points, positive is worse); TPOT: ratio to BF16 (lower is faster); Fails: number
of quality gates failed (of six).}
\label{tab:final}
{\small\setlength{\tabcolsep}{4pt}
\begin{tabular}{lrrrrrrrrrr}
\toprule
Arm & Mem & BPW & Tier & $r_{\mathrm{ko}}$ & $\Delta$acc & $\Delta_{\mathrm{num}}$ & $\Delta_{\mathrm{rea}}$ & $\Delta$fmt & TPOT & Fails \\
\midrule
Mixed NVFP4A16 + W8 (A81) & 5.917 & 6.17 & 0.931 & 0.904 & $-$0.3 & $-$0.9 & $-$0.1 & +0.14 & 0.939 & 0 \\
NVFP4 W4A4 (n4) & 5.977 & 6.25 & 0.887 & 0.821 & $-$1.7 & $-$1.3 & $-$3.9 & $-$0.18 & 0.660 & 3 \\
INT4/INT3 mix (B81) & 5.153 & 5.35 & 0.810 & 0.700 & $-$6.7 & $-$12.2 & $-$3.8 & +1.23 & 0.863 & 5 \\
GPTQ W3 (w3) & 4.864 & 5.08 & 0.748 & 0.643 & $-$10.9 & $-$20.8 & $-$8.3 & +2.92 & 0.866 & 6 \\
\bottomrule
\end{tabular}
}
\end{table}

\begin{table}[t]
\centering
\caption{FINAL-81 accuracy (\%) per gate axis for the BF16 teacher and each arm.}
\label{tab:axes}
\begin{tabular}{lrrrrrr}
\toprule
Axis & $n$ & BF16 & A81 & n4 & B81 & w3 \\
\midrule
Korean & 1,565 & 51.7 & 51.3 & 50.2 & 45.9 & 44.3 \\
English & 1,043 & 67.8 & 68.2 & 66.9 & 64.1 & 62.1 \\
Numerical & 1,320 & 90.5 & 89.5 & 89.2 & 78.3 & 69.7 \\
Reasoning & 870 & 73.9 & 73.8 & 70.0 & 70.1 & 65.6 \\
\bottomrule
\end{tabular}

\end{table}

\begin{figure}[t]
\centering
\includegraphics[width=\textwidth]{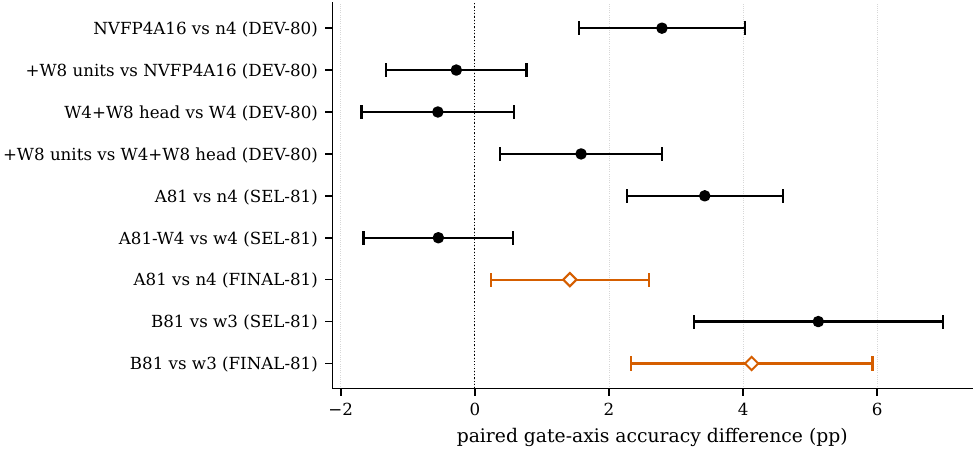}
\caption{Paired differences in gate-axis accuracy (points) with confidence intervals. DEV-80 rows use 95\%
intervals; SEL-81 and FINAL-81 rows use the Bonferroni $z$ of each track ($z=\ZA$ for A81, $z=\ZB$ for
B81). FINAL-81 rows are highlighted. The A81 advantage over n4 shrank from SEL-81 to FINAL-81 but its lower
bound stayed above zero.}
\label{fig:paired}
\end{figure}

FINAL-81 was then read once (Tables~\ref{tab:final} and~\ref{tab:axes}). \texttt{A81} passed all six
quality gates and both serving gates: memory \FNvMem{}~GiB (\FNvMemBytes{} bytes, \FNvMemMargin~GiB below
the limit), tier \FNvTier{}, Korean retention \FNvKoRet{}, overall accuracy change \FNvDacc{} points,
numerical \FNvNumDacc{} and reasoning \FNvReaDacc{}, malformed-output rate \FNvFmt{} points, and TPOT ratio
\FNvTpot. The execution check found the expected kernel on all 144 modules (\FNvKernNv{} MarlinNvFp4,
\FNvKernMarlin{} Marlin INT8) and an INT8 output head. Its Korean retention cleared the gate by only
\FNvKoMargin, a narrow pass margin given only \NFinKo{} Korean items. The reference n4 failed
three gates: Korean retention \FNfKoRet, tier \FNfTier{} and reasoning change \FNfReaDacc{} points
(Table~\ref{tab:final}). \texttt{A81} beat n4 by \FNvSup{} points \FNvSupCI{} at $z=\ZA$, so superiority was
reproduced, with a smaller effect than on SEL-81 (Figure~\ref{fig:paired}). Per axis, the gain was largest
on reasoning (\FNvSupRea{} \FNvSupReaCI{}); Korean (\FNvSupKo{} \FNvSupKoCI), English (\FNvSupEn{}
\FNvSupEnCI) and numerical (\FNvSupNum{} \FNvSupNumCI) differences were not significant on their own;
per-axis comparisons were report-only. \texttt{B81} reproduced its advantage over GPTQ W3 (\FBjSup{} points
\FBjSupCI) and failed five quality gates again.

\begin{figure}[t]
\centering
\includegraphics[width=\textwidth]{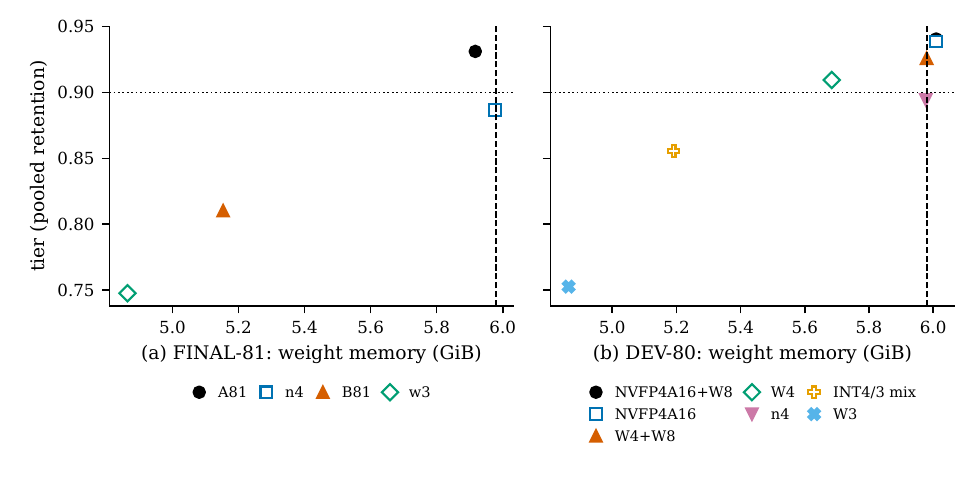}
\caption{Measured weight-loading memory against tier. (a) FINAL-81 arms. (b) DEV-80 arms. Dashed line:
memory limit (\GateMem~GiB, strict); dotted line: tier gate (\GateTier). Only A81 lies left of the memory
line and above the tier line. Both NVFP4A16 arms in (b) exceed the memory limit.}
\label{fig:memtier}
\end{figure}

\subsection{What the gain over W4A4 consists of}
\label{sec:attr}

The comparison with n4 changes two things at once: activations go from FP4 to 16 bits, and 46 modules
gain INT8 protection. The D80 development set provides a partial, exploratory decomposition
(Table~\ref{tab:paired}); it does not identify the effect of the D81 allocation on an independent set. Moving
from n4 to NVFP4 W4A16 alone raised paired accuracy by \DvNvaVsNf{} points \DvNvaVsNfCI{}; adding INT8
modules on top of it changed accuracy by \DvNvhVsNva{} points \DvNvhVsNvaCI{}, which is not significant.
These exploratory comparisons suggest that much of the gain relative to W4A4 is associated with keeping
activations at 16-bit precision. The catch is memory. NVFP4 W4A16
with a BF16 output head measured \MemNvaSixteen~GiB, above the limit. Quantizing the output head to INT8
saves about \MemLmhSave~GiB (a smoke load measured \MemNvaLmh~GiB for this combination), and the 46 INT8
modules then add \AlNvUsed~GiB, giving the final \FNvMem~GiB. The INT8 head is what makes 16-bit activations
fit, and the INT8 modules occupy the remaining headroom. We never measured the quality of NVFP4 W4A16 with
an INT8 head and no INT8 modules, nor a random choice of 46 INT8 modules, so the contribution of the
allocation itself is not identified. On the GPTQ W4 body the picture was similar: an INT8 head alone did
not help (\DvWhVsWf{} \DvWhVsWfCI{}), INT8 modules on top of it did (\DvHmVsWh{} \DvHmVsWhCI{}), and that
configuration later failed on the final set (Section~\ref{sec:devfinal}).

\begin{table}[t]
\centering
\caption{Paired gate-axis accuracy differences used for attribution (B $-$ A, points). DEV-80 rows use
$z=1.96$; SEL-81 and FINAL-81 rows use the Bonferroni $z$ of each track.}
\label{tab:paired}
\begin{tabular}{llrrr}
\toprule
Comparison (B $-$ A) & Set & $z$ & Diff (pp) & CI (pp) \\
\midrule
NVFP4A16 $-$ n4 & DEV-80 & 1.96 & +2.8 & [+1.6, +4.0] \\
+W8 units $-$ NVFP4A16 & DEV-80 & 1.96 & $-$0.3 & [$-$1.3, +0.8] \\
W4+W8 head $-$ W4 & DEV-80 & 1.96 & $-$0.6 & [$-$1.7, +0.6] \\
+W8 units $-$ W4+W8 head & DEV-80 & 1.96 & +1.6 & [+0.4, +2.8] \\
A81 $-$ n4 & SEL-81 & 2.24 & +3.4 & [+2.3, +4.6] \\
A81-W4 $-$ w4 & SEL-81 & 2.24 & $-$0.5 & [$-$1.7, +0.6] \\
A81 $-$ n4 & FINAL-81 & 2.24 & +1.4 & [+0.2, +2.6] \\
B81 $-$ w3 & SEL-81 & 2.58 & +5.1 & [+3.3, +7.0] \\
B81 $-$ w3 & FINAL-81 & 2.58 & +4.1 & [+2.3, +5.9] \\
\bottomrule
\end{tabular}

\end{table}

\subsection{Korean-heavy calibration damaged reasoning}
\label{sec:cal}

Calibration sets had 128 sequences of 2,048 tokens. The default set contained no Korean text; the
Korean-enriched sets replaced 25\% or 50\% of the rows with Korean web text. On the diagnostic set, 50\%
Korean calibration raised Korean retention of the W3 base by \DgCalKo{} points \DgCalKoCI{} and reasoning
by \DgCalRea{}, so it was chosen for the calibration test. These diagnostic intervals were exploratory
(independent-sample approximations). The preregistered paired test on SEL-81 compared \texttt{B81} with the
same allocation recalibrated on that set. The Korean difference was \CalKo{} points \CalKoCI{}, not
significant, while reasoning fell by \CalRea{} points \CalReaCI{}. The diagnostic gain did not transfer.
Calibration-language changes have to be evaluated against every required capability; a calibration
improvement for one language should not be declared from that language alone.

\subsection{Serving behaviour}
\label{sec:serving}

\begin{table}[t]
\centering
\caption{FINAL-81 serving at batch size 1 (32 prompts, 256 output tokens). Mem in GiB; TTFT in ms (p50
and p95); TPOT is the median in ms; tok/s is single-stream output throughput.}
\label{tab:serving}
\begin{tabular}{lrrrrrr}
\toprule
Arm & Mem & TTFT$_{50}$ & TTFT$_{95}$ & TPOT & tok/s & TPOT ratio \\
\midrule
BF16 teacher & 15.27 & 15.3 & 20.3 & 3.915 & 255.4 & 1.000 \\
Mixed NVFP4A16 + W8 (A81) & 5.92 & 15.4 & 16.2 & 3.676 & 272.0 & 0.939 \\
NVFP4 W4A4 (n4) & 5.98 & 42.3 & 44.0 & 2.584 & 387.0 & 0.660 \\
INT4/INT3 mix (B81) & 5.15 & 16.8 & 17.5 & 3.379 & 295.9 & 0.863 \\
GPTQ W3 (w3) & 4.86 & 16.3 & 16.9 & 3.389 & 295.0 & 0.866 \\
\bottomrule
\end{tabular}

\end{table}

Formats differ in where they are fast (Table~\ref{tab:serving}). NVFP4 W4A4 produced \FNfTok{} output
tokens per second against \TTok{} for BF16, but its median TTFT was \FNfTtft{}~ms against \TTtft{}~ms.
\texttt{A81} delivered \FNvTok{} tokens per second with a median TTFT of \FNvTtft{}~ms, close to BF16 on
both. These are batch-1 measurements on short prompts and do not generalize to other batch sizes, prompt
lengths or concurrency levels.

\section{Discussion}

\paragraph{Deployment feasibility has several independent conditions.}
An INT2 checkpoint can execute in a compact footprint and fail generation entirely. An FP4 format can lead
on output throughput and lag on first-token latency. A strong candidate can miss the memory limit by tens of
megabytes once loaded, as NVFP4 W4A16 with a BF16 head did. A configuration that leads on a development set
can fail on held-out items. Each condition has to be checked on its own.

\paragraph{Aggregate scores hide language-specific failures.}
Several configurations had acceptable aggregate or English results and failed the Korean gate. This does not
show that Korean is generally more sensitive to quantization; it shows that per-language requirements can
decide whether a compressed model is acceptable for a given service, and that they must be set before
candidates are chosen.

\paragraph{More precision on selected modules is not reliably better.}
The D80 INT4/INT8 result, the non-additivity of restoration and the weak INT8 contribution on the NVFP4
body all point the same way. Which modules to protect depends on interactions, the task mix, the activation
path and the supported formats; a ranking from isolated measurements is a weak guide.

\paragraph{Runtime support shapes the search space.}
Fused-module boundaries, kernel availability, loading overhead and activation precision determine which
mixtures can actually run. Among the configurations we tested, a feasible path was to retain 16-bit
activations and use an INT8 output head to create memory headroom; no global cost-optimality claim follows.
Including such
constraints in the search changes which configuration wins.

\section{Limitations}

\paragraph{One model, one GPU.} All results are for Qwen3-8B on B200 with vLLM 0.29.0. Other models, scales,
architectures and runtimes may behave differently.

\paragraph{Attribution is incomplete.} The gain of \texttt{A81} over n4 combines activation precision and
INT8 protection. We have neither a quality measurement of NVFP4 W4A16 with an INT8 head and no INT8 modules
nor a random-allocation control, and we compared against no published allocation method.

\paragraph{Small margins.} \texttt{A81}'s Korean retention passed its gate by \FNvKoMargin, and its
superiority lower bound was \FNvSupLo{} points. A modest shift in the item mix could change either verdict.

\paragraph{Effect shrinkage.} The advantage over n4 fell from \SNvSup{} points on SEL-81 to \FNvSup{} on
FINAL-81, consistent with some selection optimism even under a frozen protocol.

\paragraph{Checkpoints not preserved.} Final-stage checkpoints were built, fingerprinted and served inside
the evaluation job and not retained. The SEL-81 and FINAL-81 builds of \texttt{A81}, from the same recipe,
calibration and code, have different SHA-256 digests, so rebuilds are not bitwise reproducible. We have not
measured how much quality varies between rebuilds.

\paragraph{Allocator prediction.} The allocator predicted \AlNvPred~GiB against an internal planning limit
of \AlNvLimit~GiB, leaving \AlNvHead~MiB; the measured counter was \FNvMem~GiB. The prediction was
conservative here, but it is a planning aid, not a guarantee.

\paragraph{Underpowered track B.} Track B was preregistered as underpowered; its null results are not
evidence of no effect.

\paragraph{Serving scope.} Latency was measured at batch size 1 with short prompts. Throughput under
concurrency, long contexts, KV-cache capacity, peak memory and energy were not measured.

\paragraph{Study-specific gates.} The thresholds are requirements of this study, not industry standards.

\paragraph{No release.} Passing the preregistered gates is not a product release. No checkpoint accompanies
this paper.

\section{Future work}

The most important next step is a causal ablation under one protocol on a fresh held-out set: NVFP4 W4A4,
NVFP4 W4A16 with an INT8 head and no INT8 modules, the same with 46 randomly chosen INT8 modules, and the
ranked allocation, all at matched memory. Interaction-aware allocators \citep{zhao2025impq,zheng2026mixllm}
should be compared under the same runtime constraints. Repeating the study on other model families and
sizes, measuring rebuild variance, and extending the serving study to concurrency, long contexts and memory
headroom would show how far these findings extend.

\section{Conclusion}

We studied low-bit and mixed-precision quantization of Qwen3-8B on NVIDIA B200 under execution, memory,
latency and multilingual quality constraints. Native INT2 and INT3 execution did not ensure usable output,
isolated module sensitivity was a poor guide to mixtures, a development-set winner failed on held-out
items, and Korean-heavy calibration damaged reasoning. A configuration with 16-bit activations, NVFP4
weights, 46 INT8 modules and an INT8 output head passed all preregistered gates on an untouched final set,
with \FNvMem~GiB of weight-loading memory, Korean retention \FNvKoRet{} and a TPOT ratio of \FNvTpot, and
outperformed NVFP4 W4A4 by \FNvSup{} points \FNvSupCI. Exploratory development-set comparisons suggest
activation precision is the larger contributor, but the effect of the selected INT8 allocation remains
unidentified. Quantization for deployment is a joint numerical and systems
problem, and the measurement protocol, including what it rejects, matters as much as the format.

\bibliographystyle{plainnat}
\bibliography{refs}

\begin{thebibliography}{45}
\providecommand{\natexlab}[1]{#1}
\providecommand{\url}[1]{\texttt{#1}}
\expandafter\ifx\csname urlstyle\endcsname\relax
  \providecommand{\doi}[1]{doi: #1}\else
  \providecommand{\doi}{doi: \begingroup \urlstyle{rm}\Url}\fi

\bibitem[{Ant Group and the vLLM Community}(2026)]{humming2026}
{Ant Group and the vLLM Community}.
\newblock Announcing {Humming}: Fast, flexible kernels for quantized inference.
\newblock vLLM Blog,
  \url{https://vllm-project.github.io/2026/10/19/humming.html}, 2026.

\bibitem[Ashkboos et~al.(2024)Ashkboos, Mohtashami, Croci, Li, Cameron, Jaggi,
  Alistarh, Hoefler, and Hensman]{ashkboos2024quarot}
Saleh Ashkboos, Amirkeivan Mohtashami, Maximilian~L. Croci, Bo~Li, Pashmina
  Cameron, Martin Jaggi, Dan Alistarh, Torsten Hoefler, and James Hensman.
\newblock {QuaRot}: Outlier-free 4-bit inference in rotated {LLMs}.
\newblock In \emph{Advances in Neural Information Processing Systems
  (NeurIPS)}, 2024.

\bibitem[Chee et~al.(2023)Chee, Cai, Kuleshov, and Sa]{chee2023quip}
Jerry Chee, Yaohui Cai, Volodymyr Kuleshov, and Christopher~De Sa.
\newblock {QuIP}: 2-bit quantization of large language models with guarantees.
\newblock In \emph{Advances in Neural Information Processing Systems
  (NeurIPS)}, 2023.

\bibitem[Chen et~al.(2025)Chen, Shao, Xu, Wang, Gao, Zhang, and
  Luo]{chen2025efficientqat}
Mengzhao Chen, Wenqi Shao, Peng Xu, Jiahao Wang, Peng Gao, Kaipeng Zhang, and
  Ping Luo.
\newblock {EfficientQAT}: Efficient quantization-aware training for large
  language models.
\newblock In \emph{Proceedings of the 63rd Annual Meeting of the Association
  for Computational Linguistics (ACL)}, 2025.

\bibitem[Chimoto et~al.(2026)Chimoto, Elhoushi, and
  Bassett]{chimoto2026calibrating}
Everlyn~Asiko Chimoto, Mostafa Elhoushi, and Bruce Bassett.
\newblock Calibrating beyond {English}: Language diversity for better quantized
  multilingual {LLMs}.
\newblock In \emph{Proceedings of the 19th Conference of the European Chapter
  of the Association for Computational Linguistics (EACL)}, 2026.
\newblock 2026.eacl-long.223.

\bibitem[Clark et~al.(2018)Clark, Cowhey, Etzioni, Khot, Sabharwal, Schoenick,
  and Tafjord]{clark2018arc}
Peter Clark, Isaac Cowhey, Oren Etzioni, Tushar Khot, Ashish Sabharwal, Carissa
  Schoenick, and Oyvind Tafjord.
\newblock Think you have solved question answering? try {ARC}, the {AI2}
  reasoning challenge.
\newblock \emph{arXiv preprint arXiv:1803.05457}, 2018.

\bibitem[Dettmers et~al.(2022)Dettmers, Lewis, Belkada, and
  Zettlemoyer]{dettmers2022llmint8}
Tim Dettmers, Mike Lewis, Younes Belkada, and Luke Zettlemoyer.
\newblock {LLM.int8()}: 8-bit matrix multiplication for transformers at scale.
\newblock In \emph{Advances in Neural Information Processing Systems
  (NeurIPS)}, 2022.
\newblock arXiv:2208.07339.

\bibitem[Dettmers et~al.(2023)Dettmers, Pagnoni, Holtzman, and
  Zettlemoyer]{dettmers2023qlora}
Tim Dettmers, Artidoro Pagnoni, Ari Holtzman, and Luke Zettlemoyer.
\newblock {QLoRA}: Efficient finetuning of quantized {LLMs}.
\newblock In \emph{Advances in Neural Information Processing Systems
  (NeurIPS)}, 2023.

\bibitem[Dettmers et~al.(2024)Dettmers, Svirschevski, Egiazarian, Kuznedelev,
  Frantar, Ashkboos, Borzunov, Hoefler, and Alistarh]{dettmers2024spqr}
Tim Dettmers, Ruslan Svirschevski, Vage Egiazarian, Denis Kuznedelev, Elias
  Frantar, Saleh Ashkboos, Alexander Borzunov, Torsten Hoefler, and Dan
  Alistarh.
\newblock {SpQR}: A sparse-quantized representation for near-lossless {LLM}
  weight compression.
\newblock In \emph{International Conference on Learning Representations
  (ICLR)}, 2024.

\bibitem[Dong et~al.(2019)Dong, Yao, Gholami, Mahoney, and
  Keutzer]{dong2019hawq}
Zhen Dong, Zhewei Yao, Amir Gholami, Michael~W. Mahoney, and Kurt Keutzer.
\newblock {HAWQ}: Hessian {AWare} quantization of neural networks with
  mixed-precision.
\newblock In \emph{Proceedings of the IEEE/CVF International Conference on
  Computer Vision (ICCV)}, 2019.

\bibitem[Dong et~al.(2020)Dong, Yao, Arfeen, Gholami, Mahoney, and
  Keutzer]{dong2020hawqv2}
Zhen Dong, Zhewei Yao, Daiyaan Arfeen, Amir Gholami, Michael~W. Mahoney, and
  Kurt Keutzer.
\newblock {HAWQ-V2}: Hessian aware trace-weighted quantization of neural
  networks.
\newblock In \emph{Advances in Neural Information Processing Systems
  (NeurIPS)}, 2020.

\bibitem[Egiazarian et~al.(2024)Egiazarian, Panferov, Kuznedelev, Frantar,
  Babenko, and Alistarh]{egiazarian2024aqlm}
Vage Egiazarian, Andrei Panferov, Denis Kuznedelev, Elias Frantar, Artem
  Babenko, and Dan Alistarh.
\newblock Extreme compression of large language models via additive
  quantization.
\newblock In \emph{Proceedings of the 41st International Conference on Machine
  Learning (ICML)}, 2024.

\bibitem[Egiazarian et~al.(2026)]{egiazarian2026fp4}
Vage Egiazarian et~al.
\newblock Bridging the gap between promise and performance for microscaling
  {FP4} quantization.
\newblock In \emph{International Conference on Learning Representations
  (ICLR)}, 2026.

\bibitem[Frantar et~al.(2023)Frantar, Ashkboos, Hoefler, and
  Alistarh]{frantar2023gptq}
Elias Frantar, Saleh Ashkboos, Torsten Hoefler, and Dan Alistarh.
\newblock {GPTQ}: Accurate post-training quantization for generative
  pre-trained transformers.
\newblock In \emph{International Conference on Learning Representations
  (ICLR)}, 2023.
\newblock arXiv:2210.17323.

\bibitem[Frantar et~al.(2024)Frantar, Castro, Chen, Hoefler, and
  Alistarh]{frantar2024marlin}
Elias Frantar, Roberto~L. Castro, Jiale Chen, Torsten Hoefler, and Dan
  Alistarh.
\newblock {MARLIN}: Mixed-precision auto-regressive parallel inference on large
  language models.
\newblock \emph{arXiv preprint arXiv:2408.11743}, 2024.

\bibitem[Hendrycks et~al.(2021)Hendrycks, Burns, Basart, Zou, Mazeika, Song,
  and Steinhardt]{hendrycks2021mmlu}
Dan Hendrycks, Collin Burns, Steven Basart, Andy Zou, Mantas Mazeika, Dawn
  Song, and Jacob Steinhardt.
\newblock Measuring massive multitask language understanding.
\newblock In \emph{International Conference on Learning Representations
  (ICLR)}, 2021.

\bibitem[Jang et~al.(2022)Jang, Kim, Kwon, and Davis]{jang2022kobest}
Myeongjun Jang, Dohyung Kim, Deuk~Sin Kwon, and Eric Davis.
\newblock {KoBEST}: {Korean} balanced evaluation of significant tasks.
\newblock In \emph{Proceedings of the 29th International Conference on
  Computational Linguistics (COLING)}, 2022.

\bibitem[Kim et~al.(2024)Kim, Hooper, Gholami, Dong, Li, Shen, Mahoney, and
  Keutzer]{kim2024squeezellm}
Sehoon Kim, Coleman Hooper, Amir Gholami, Zhen Dong, Xiuyu Li, Sheng Shen,
  Michael~W. Mahoney, and Kurt Keutzer.
\newblock {SqueezeLLM}: Dense-and-sparse quantization.
\newblock In \emph{Proceedings of the 41st International Conference on Machine
  Learning (ICML)}, 2024.

\bibitem[Ko et~al.(2025)Ko, Son, and Choi]{ko2025hrm8k}
Hyunwoo Ko, Guijin Son, and Dasol Choi.
\newblock Understand, solve and translate: Bridging the multilingual
  mathematical reasoning gap.
\newblock \emph{arXiv preprint arXiv:2501.02448}, 2025.

\bibitem[Kwon et~al.(2023)Kwon, Li, Zhuang, Sheng, Zheng, Yu, Gonzalez, Zhang,
  and Stoica]{kwon2023pagedattention}
Woosuk Kwon, Zhuohan Li, Siyuan Zhuang, Ying Sheng, Lianmin Zheng, Cody~Hao Yu,
  Joseph~E. Gonzalez, Hao Zhang, and Ion Stoica.
\newblock Efficient memory management for large language model serving with
  {PagedAttention}.
\newblock In \emph{Proceedings of the 29th ACM Symposium on Operating Systems
  Principles (SOSP)}, 2023.
\newblock \doi{10.1145/3600006.3613165}.

\bibitem[Lin et~al.(2024{\natexlab{a}})Lin, Xu, Wu, Cui, Zhang, Mou, Song, Sun,
  and Wei]{lin2024duquant}
Haokun Lin, Haobo Xu, Yichen Wu, Jingzhi Cui, Yingtao Zhang, Linzhan Mou, Linqi
  Song, Zhenan Sun, and Ying Wei.
\newblock {DuQuant}: Distributing outliers via dual transformation makes
  stronger quantized {LLMs}.
\newblock In \emph{Advances in Neural Information Processing Systems
  (NeurIPS)}, 2024{\natexlab{a}}.

\bibitem[Lin et~al.(2024{\natexlab{b}})Lin, Tang, Tang, Yang, Chen, Wang, Xiao,
  Dang, Gan, and Han]{lin2024awq}
Ji~Lin, Jiaming Tang, Haotian Tang, Shang Yang, Wei-Ming Chen, Wei-Chen Wang,
  Guangxuan Xiao, Xingyu Dang, Chuang Gan, and Song Han.
\newblock {AWQ}: Activation-aware weight quantization for on-device {LLM}
  compression and acceleration.
\newblock In \emph{Proceedings of Machine Learning and Systems (MLSys)},
  2024{\natexlab{b}}.

\bibitem[Lin et~al.(2025)Lin, Tang, Yang, Zhang, Xiao, Gan, and
  Han]{lin2025qserve}
Yujun Lin, Haotian Tang, Shang Yang, Zhekai Zhang, Guangxuan Xiao, Chuang Gan,
  and Song Han.
\newblock {QServe}: {W4A8KV4} quantization and system co-design for efficient
  {LLM} serving.
\newblock In \emph{Proceedings of Machine Learning and Systems (MLSys)}, 2025.

\bibitem[Liu et~al.(2024{\natexlab{a}})Liu, Oguz, Zhao, Chang, Stock, Mehdad,
  Shi, Krishnamoorthi, and Chandra]{liu2024llmqat}
Zechun Liu, Barlas Oguz, Changsheng Zhao, Ernie Chang, Pierre Stock, Yashar
  Mehdad, Yangyang Shi, Raghuraman Krishnamoorthi, and Vikas Chandra.
\newblock {LLM-QAT}: Data-free quantization aware training for large language
  models.
\newblock In \emph{Findings of the Association for Computational Linguistics:
  ACL 2024}, 2024{\natexlab{a}}.

\bibitem[Liu et~al.(2025{\natexlab{a}})Liu, Zhao, Fedorov, Soran, Choudhary,
  Krishnamoorthi, Chandra, Tian, and Blankevoort]{liu2025spinquant}
Zechun Liu, Changsheng Zhao, Igor Fedorov, Bilge Soran, Dhruv Choudhary,
  Raghuraman Krishnamoorthi, Vikas Chandra, Yuandong Tian, and Tijmen
  Blankevoort.
\newblock {SpinQuant}: {LLM} quantization with learned rotations.
\newblock In \emph{International Conference on Learning Representations
  (ICLR)}, 2025{\natexlab{a}}.

\bibitem[Liu et~al.(2025{\natexlab{b}})Liu, Zhao, Huang, Chen, Zhang, Zhao,
  Roy, Jin, Xiong, Shi, Xiao, Tian, Soran, Krishnamoorthi, Blankevoort, and
  Chandra]{liu2025paretoq}
Zechun Liu, Changsheng Zhao, Hanxian Huang, Sijia Chen, Jing Zhang, Jiawei
  Zhao, Scott Roy, Lisa Jin, Yunyang Xiong, Yangyang Shi, Lin Xiao, Yuandong
  Tian, Bilge Soran, Raghuraman Krishnamoorthi, Tijmen Blankevoort, and Vikas
  Chandra.
\newblock {ParetoQ}: Scaling laws in extremely low-bit {LLM} quantization.
\newblock \emph{arXiv preprint arXiv:2502.02631}, 2025{\natexlab{b}}.

\bibitem[Liu et~al.(2024{\natexlab{b}})Liu, Yuan, Jin, Zhong, Xu, Braverman,
  Chen, and Hu]{liu2024kivi}
Zirui Liu, Jiayi Yuan, Hongye Jin, Shaochen Zhong, Zhaozhuo Xu, Vladimir
  Braverman, Beidi Chen, and Xia Hu.
\newblock {KIVI}: A tuning-free asymmetric 2bit quantization for {KV} cache.
\newblock In \emph{Proceedings of the 41st International Conference on Machine
  Learning (ICML)}, 2024{\natexlab{b}}.

\bibitem[Nagel et~al.(2020)Nagel, Amjad, van Baalen, Louizos, and
  Blankevoort]{nagel2020adaround}
Markus Nagel, Rana~Ali Amjad, Mart van Baalen, Christos Louizos, and Tijmen
  Blankevoort.
\newblock Up or down? {Adaptive} rounding for post-training quantization.
\newblock In \emph{Proceedings of the 37th International Conference on Machine
  Learning (ICML)}, 2020.

\bibitem[{Qwen Team}(2025)]{qwen3}
{Qwen Team}.
\newblock Qwen3 technical report.
\newblock \emph{arXiv preprint arXiv:2505.09388}, 2025.

\bibitem[Shao et~al.(2024)Shao, Chen, Zhang, Xu, Zhao, Li, Zhang, Gao, Qiao,
  and Luo]{shao2024omniquant}
Wenqi Shao, Mengzhao Chen, Zhaoyang Zhang, Peng Xu, Lirui Zhao, Zhiqian Li,
  Kaipeng Zhang, Peng Gao, Yu~Qiao, and Ping Luo.
\newblock {OmniQuant}: Omnidirectionally calibrated quantization for large
  language models.
\newblock In \emph{International Conference on Learning Representations
  (ICLR)}, 2024.

\bibitem[Son et~al.(2025)Son, Lee, Kim, Kim, Muennighoff, Choi, Park, Yoo, and
  Biderman]{son2025kmmlu}
Guijin Son, Hanwool Lee, Sungdong Kim, Seungone Kim, Niklas Muennighoff,
  Taekyoon Choi, Cheonbok Park, Kang~Min Yoo, and Stella Biderman.
\newblock {KMMLU}: Measuring massive multitask language understanding in
  {Korean}.
\newblock In \emph{Proceedings of the 2025 Conference of the Nations of the
  Americas Chapter of the Association for Computational Linguistics (NAACL)},
  2025.

\bibitem[Sprague et~al.(2024)Sprague, Ye, Bostrom, Chaudhuri, and
  Durrett]{sprague2024musr}
Zayne Sprague, Xi~Ye, Kaj Bostrom, Swarat Chaudhuri, and Greg Durrett.
\newblock {MuSR}: Testing the limits of chain-of-thought with multistep soft
  reasoning.
\newblock In \emph{International Conference on Learning Representations
  (ICLR)}, 2024.

\bibitem[Srivastava et~al.(2023)]{srivastava2023bigbench}
Aarohi Srivastava et~al.
\newblock Beyond the imitation game: Quantifying and extrapolating the
  capabilities of language models.
\newblock \emph{Transactions on Machine Learning Research}, 2023.

\bibitem[Sun et~al.(2025)Sun, Liu, Bai, Bao, Zhao, Li, Hu, Yu, Hou, Yuan,
  Jiang, Liu, and Yao]{sun2025flatquant}
Yuxuan Sun, Ruikang Liu, Haoli Bai, Han Bao, Kang Zhao, Yuening Li, Jiaxin Hu,
  Xianzhi Yu, Lu~Hou, Chun Yuan, Xin Jiang, Wulong Liu, and Jun Yao.
\newblock {FlatQuant}: Flatness matters for {LLM} quantization.
\newblock In \emph{Proceedings of the 42nd International Conference on Machine
  Learning (ICML)}, 2025.

\bibitem[Suzgun et~al.(2023)Suzgun, Scales, Sch{\"a}rli, Gehrmann, Tay, Chung,
  Chowdhery, Le, Chi, Zhou, and Wei]{suzgun2023bbh}
Mirac Suzgun, Nathan Scales, Nathanael Sch{\"a}rli, Sebastian Gehrmann, Yi~Tay,
  Hyung~Won Chung, Aakanksha Chowdhery, Quoc~V. Le, Ed~H. Chi, Denny Zhou, and
  Jason Wei.
\newblock Challenging {BIG-Bench} tasks and whether chain-of-thought can solve
  them.
\newblock In \emph{Findings of the Association for Computational Linguistics:
  ACL 2023}, 2023.

\bibitem[Tseng et~al.(2024)Tseng, Chee, Sun, Kuleshov, and
  Sa]{tseng2024quipsharp}
Albert Tseng, Jerry Chee, Qingyao Sun, Volodymyr Kuleshov, and Christopher~De
  Sa.
\newblock {QuIP\#}: Even better {LLM} quantization with {Hadamard} incoherence
  and lattice codebooks.
\newblock In \emph{Proceedings of the 41st International Conference on Machine
  Learning (ICML)}, 2024.

\bibitem[{vLLM Project}(2026)]{vllm029}
{vLLM Project}.
\newblock {vLLM} v0.29.0 documentation.
\newblock \url{https://docs.vllm.ai/en/v0.29.0/}, 2026.

\bibitem[Waskito et~al.(2026)]{waskito2026multilinguality}
Muhammad Rafif~Priyo Waskito et~al.
\newblock Multilinguality after quantization: Low-bits, calibration languages,
  mixed-precisions, and cross-linguality.
\newblock In \emph{Findings of the Association for Computational Linguistics:
  IJCNLP-AACL}, 2026.
\newblock 2026.findings-aacl.9.

\bibitem[Xiao et~al.(2023)Xiao, Lin, Seznec, Wu, Demouth, and
  Han]{xiao2023smoothquant}
Guangxuan Xiao, Ji~Lin, Mickael Seznec, Hao Wu, Julien Demouth, and Song Han.
\newblock {SmoothQuant}: Accurate and efficient post-training quantization for
  large language models.
\newblock In \emph{Proceedings of the 40th International Conference on Machine
  Learning (ICML)}, 2023.

\bibitem[Yao et~al.(2022)Yao, Aminabadi, Zhang, Wu, Li, and
  He]{yao2022zeroquant}
Zhewei Yao, Reza~Yazdani Aminabadi, Minjia Zhang, Xiaoxia Wu, Conglong Li, and
  Yuxiong He.
\newblock {ZeroQuant}: Efficient and affordable post-training quantization for
  large-scale transformers.
\newblock In \emph{Advances in Neural Information Processing Systems
  (NeurIPS)}, 2022.

\bibitem[Ye et~al.(2025)Ye, Chen, Lai, Lin, Zhang, Wang, Chen, Kasikci, Grover,
  Krishnamurthy, and Ceze]{ye2025flashinfer}
Zihao Ye, Lequn Chen, Ruihang Lai, Wuwei Lin, Yineng Zhang, Stephanie Wang,
  Tianqi Chen, Baris Kasikci, Vinod Grover, Arvind Krishnamurthy, and Luis
  Ceze.
\newblock {FlashInfer}: Efficient and customizable attention engine for {LLM}
  inference serving.
\newblock In \emph{Proceedings of Machine Learning and Systems (MLSys)}, 2025.

\bibitem[Zellers et~al.(2019)Zellers, Holtzman, Bisk, Farhadi, and
  Choi]{zellers2019hellaswag}
Rowan Zellers, Ari Holtzman, Yonatan Bisk, Ali Farhadi, and Yejin Choi.
\newblock {HellaSwag}: Can a machine really finish your sentence?
\newblock In \emph{Proceedings of the 57th Annual Meeting of the Association
  for Computational Linguistics (ACL)}, 2019.

\bibitem[Zhao et~al.(2025)]{zhao2025impq}
Junchen Zhao et~al.
\newblock {IMPQ}: Interaction-aware layerwise mixed precision quantization for
  {LLMs}.
\newblock \emph{arXiv preprint arXiv:2509.15455}, 2025.

\bibitem[Zhao et~al.(2024)Zhao, Lin, Zhu, Ye, Chen, Zheng, Ceze, Krishnamurthy,
  Chen, and Kasikci]{zhao2024atom}
Yilong Zhao, Chien-Yu Lin, Kan Zhu, Zihao Ye, Lequn Chen, Size Zheng, Luis
  Ceze, Arvind Krishnamurthy, Tianqi Chen, and Baris Kasikci.
\newblock Atom: Low-bit quantization for efficient and accurate {LLM} serving.
\newblock In \emph{Proceedings of Machine Learning and Systems (MLSys)}, 2024.

\bibitem[Zheng et~al.(2026)Zheng, Song, and Liu]{zheng2026mixllm}
Zhen Zheng, Xiaonan Song, and Chuanjie Liu.
\newblock {MixLLM}: {LLM} quantization with global mixed-precision between
  output-features and highly-efficient system design.
\newblock In \emph{Proceedings of Machine Learning and Systems (MLSys)}, 2026.

\end{thebibliography}

\appendix
\FloatBarrier
\section{FINAL-81 details}

FINAL-81 has SHA-256 prefix \texttt{\FinSetSha} and \NFin{} items. It was read once for the BF16 teacher,
the two track references (n4 and GPTQ W3), \texttt{A81} and \texttt{B81}, under the frozen protocol. The
\texttt{A81} build served in this job used \FNvMemBytes{} bytes of weight-loading memory; its checkpoint
of \NvCkptBytes{} bytes had SHA-256 prefix \texttt{\FinSha}, while the SEL-81 build had \texttt{\SelSha}.
All four quantized arms passed the execution check with zero kernel mismatches; \texttt{B81} ran
\FBjKernHum{} modules through Humming (INT3) and \FBjKernMarlin{} through Marlin (INT4). The deployment
verdict for \texttt{A81} was \texttt{\FinVerdict}; track B had no deployment candidate. Per-arm results are
in Tables~\ref{tab:final}--\ref{tab:serving}.

\section{Ablations needed for an allocation claim}

The present results do not isolate the allocation procedure. At minimum, the following arms are needed on
a fresh held-out set: NVFP4 W4A4 (reference); NVFP4 W4A16 with an INT8 head and no INT8 modules (activation
precision only); the same with 46 randomly chosen INT8 modules (protection budget only); the same with the
46 ranked modules (allocation); and a published allocation method at matched measured memory. Where memory
differs, both accuracy at matched memory and the full measured quality--memory frontier should be reported.
Items consumed by FINAL-81 must not be reused for tuning.

\section{Reproducibility}

For each stage we keep the set manifests and hashes, the pinned model revision, the quantization recipes and
calibration hashes, the frozen ranking and allocation code with its output hashes, the per-module precision
assignments, environment manifests, raw memory counters, kernel and weight-type verification records,
decoding settings and the preregistration records. Every number in this paper is generated from those
ledgers by one script. Final-stage checkpoints were not preserved (see Limitations). Infrastructure
identifiers and credentials are not part of the record.

\end{document}